\documentclass[12pt,preprint]{aastex}
\usepackage{graphicx}
\def\gtrsim{\mathrel{\hbox{\rlap{\hbox{\lower3pt\hbox{$\sim$}}}\raise2pt\hbox{$>$}}}}
\def\lesssim{\mathrel{\hbox{\rlap{\hbox{\lower3pt\hbox{$\sim$}}}\raise2pt\hbox{$<$}}}}

\newcommand{\gae}{\lower 2pt \hbox{$\, \buildrel {\scriptstyle >}\over {\scriptstyle
\sim}\,$}}
\newcommand{\lae}{\lower 2pt \hbox{$\, \buildrel {\scriptstyle <}\over {\scriptstyle
\sim}\,$}}

\usepackage{amsmath}
\usepackage{color} 
\usepackage{ulem} 

\begin{document}

\title{Binary Pulsar J0737-3039 -- Evidence for a new core collapse and
  neutron star formation mechanism}

\author{Simone Dall'Osso\altaffilmark{1a}, Tsvi Piran\altaffilmark{1b} \& Nir Shaviv\altaffilmark{1c}}
\altaffiltext{1}{Racah Institute for Physics, The Hebrew University, Jerusalem, 91904, Israel}
\email{(a) simone@phys.huji.ac.il; (b) tsvi.piran@mail.huji.ac.il; (c) shaviv@phys.huji.ac.il }

\begin{abstract}
The binary pulsar J0737-3039 is 
the only known system having two observable pulsars, thus offering a
unique laboratory to test general relativity and explore pulsar
  physics. Based on the low eccentricity and the position within the
galactic plane, \cite{PiranShavivAstroPh,PiranShavivPRL} argued that pulsar B had a
non-standard formation scenario 
with little or no mass ejection. 
They 
have also predicted that the system would have a very slow proper motion. Pulsar
timing measurements   \citep{Kramer2006,Deller2009} confirmed this prediction.
The recent observations of  the alignment between the spin of pulsar A and the 
binary orbit is also in agreement with this scenario. 
Detailed simulations of the formation process of pulsar B enable us to show that 
 its progenitor, just before the 
collapse, was a massive O-Ne-Mg white dwarf surrounded  by a tenuous, 0.1-0.16 M$_\odot$, envelope. This envelope was
 ejected when the white dwarf collapsed to form a neutron star.  Pulsar B was born as a slow rotator (spin period $\sim 
1 $ s) and a kick received when the pulsar formed changed its spin direction to the current one.    
This realization  sheds light on the angular momentum 
evolution of the progenitor star, a process which is strongly affected by 
interaction with the binary companion.
The slow proper motion of the system 
also implies that the system must have undergone a phase of mass transfer in which 
Star A shed a significant fraction of its mass onto B.

\end{abstract}
\section{Introduction}

The binary system PSR J0737 contains a millisecond pulsar (Pulsar A), with a
spin of $\approx 23$ ms, and a regular pulsar (Pulsar B) with a spin period of
$\approx 2.7$ s. Following its discovery, it was suggested that the progenitor
of pulsar B 
must have  been a $\sim (2.1 \div 2.3) M_\odot$ He star   \citep{Dewi2004,Willems2004}. The large mass
loss in the SN leading to the birth of the pulsar would have given a large
kick, $\sim$ 150 km/s, keeping the system in a low eccentricity
orbit. However, \cite{PiranShavivAstroPh,PiranShavivPRL} used the small orbital eccentricity of the system and the small 
height from the galactic plane, to infer that pulsar B must have formed with 
a  small mass ejection (less than 0.1 M$_{\odot}$) and no kick
velocity.   They have
predicted  that the system should have a very low
proper motion, $v_{\rm cm,\perp} < 15$~km/s\footnote{The
    subscript $\perp$ indicates throughout components 
of the velocity orthogonal to our line of sight.}. This was later
confirmed through pulsar timing measurements   \citep{Kramer2006,Deller2009},
showing that the system has
a proper motion of $10 \pm 1$ km/s. This value 
is consistent with the random stellar motion in the galaxy and thus provides an 
upper limit on the actual kick velocity, $\Delta v_{\rm cm, \perp}$,
obtained during the collapse. 

The confirmed low proper motion implies that pulsar B must have formed with
little mass ejection, making it the first neutron star (NS) known to have been formed
in the collapse of an unstable white dwarf (WD). This could not have been through accretion
induced collapse, because its companion was already a NS and could not
  provide matter for accretion.
A WD dwarf could become unstable either through cooling, if it was
  above the Chandrasekhar mass, or, as suggested four decades
ago  \citep{Finzi}, through $e$-capture changing the average number of 
electrons in an O-Ne-Mg WD.

Besides being the best laboratory for general relativity  \citep{Kramer2006}, 
and offering the first evidence for a new channel for neutron star formation,
more can be learned, from this system, on its earlier evolutionary phases. 
Remarkably, the slow proper motion of the system  also 
 constraints the mass ejection that took place when the first pulsar, Pulsar A, formed. When combined with the 
unique measurement of the spin direction of both
pulsars it enables us  us to clarify some fundamental issues in the evolutionary
path of the progenitor system.

We use the following dynamic parameters of the system \citep{Kramer2006}: 
the pulsars's masses $M_A= 1.34$ M$_{\odot}$ and $M_B= 1.25$ M$_{\odot}$ of the pulsars, 
the eccentricity of the orbit, $e = 0.0878$, its period $P=0.102$ d ($\simeq 
2.45$ hr) and the C.M. proper motion  $v_\perp = (10\pm 1)$ km/s. The height
of the system above the galactic plane is $h= 50$pc \citep{PiranShavivAstroPh}. 
We also use the age of pulsar B, estimated to be $\le 50$ Myr, the long spin  
period (2.7 sec) of Pulsar B \citep{Kramer2006} and the inclinations of the 
spins of pulsar A and B relative to the orbital angular momentum, 
$\simeq 3.2^{\circ}$ \citep{Ferd13} and $\sim 130^{\circ}$ \citep{LyTh05,
  Bre08, Farr11}, respectively. We find it remarkable how much information one 
can obtain on the evolution of the system just from these  observables.

We begin in \S \ref{pulsarB}  using the kinematics of the system to limit the mass of pulsar B's progenitor and from this 
to obtain limits on its progenitor's structure and on the collapse that formed the NS. In \S \ref{pulsarA} we show that the slow proper motion also limits the mass of pulsar A's progenitor just before the first SN took place, to be smaller than the mass of pulsar B's progenitor at that time. As initially A must have been more massive this provides direct evidence for a significant mass loss of the progenitor of pulsar A. In \S \ref{masstrasfer} we turn to the late mass transfer between pulsar B's progenitor and pulsar A, showing that this mass transfer is consistent with the spin and the mass of this pulsar. Finally, in \S  \ref{spinB} we examine the orientations of the spins of the two pulsars. We conclude and summarize our results in \S \ref{conclusions}.

\section{Pulsar B's progenitor}
\label{pulsarB}

\subsection{Kinematic considerations: limits on the progenitor's mass}

We begin by obtaining the probability distribution function for the progenitor mass and kick velocity given the additional constraints we know today about the proper motion of the system and the alignment between pulsar A 's spin and the orbital angular momentum. Thus, we repeat our previous analysis  \citep{PiranShavivPRL}, obtaining more stringent limits. 

\begin{figure}[t]
\centerline{
\includegraphics[width=10.7 cm]{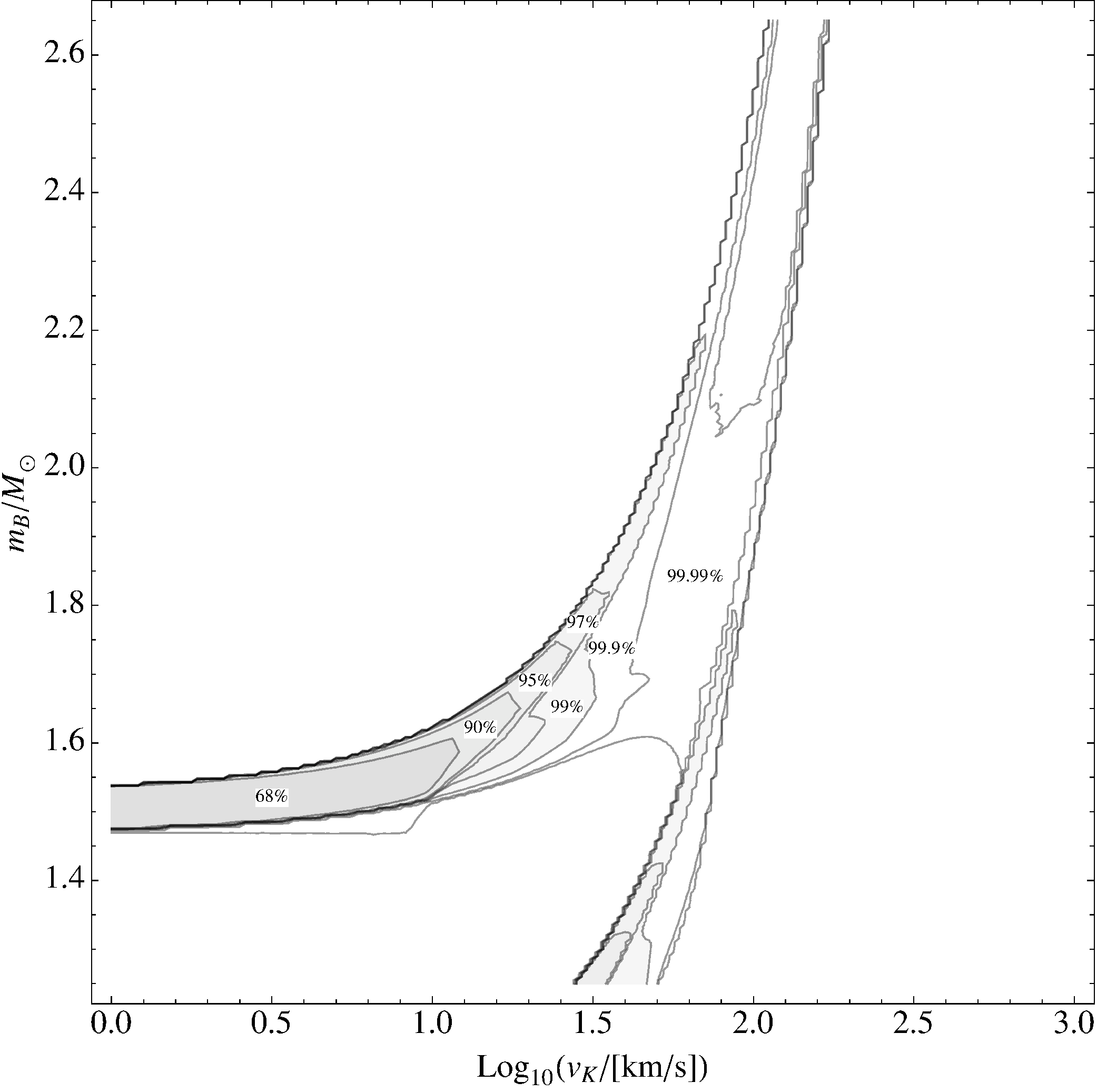}}
\caption{The probability that a binary system will end up as observed, given a
  constant prior in progenitor mass and in $\log({\rm v}_{\rm kick})$. 
  This probability is obtained by  imposing that (i) the binary should
  be within 50pc from the galactic plane; (ii)  with an eccentricity between
  0.088 and 0.14, given that 50 Myrs before, the progenitor system had a
  circular orbit, (iii)  star B had a progenitor mass $m_{B}$, and it obtained
  a random kick velocity of size $v_{kick}$; (iv) the  
  observed proper motion today should be $\le 10 \pm 1$ km/s; (v)
  the pulsar A / orbit alignment of 3.7$^o$. 
  The standard core-collapse solution of a He star requires $m_{B}$ larger than about $2.1M_{\odot}$. With a fine tuned kick velocity, this type of a solution is ruled out at the 99\% c.l. Solutions with $m_{B}$ between 1.47 and $1.53~M_{\odot}$ (and kick
    velocities smaller than $\sim 20$ km/s) are kinematically more favorable. 
Lower mass solutions with high kick velocities are kinematically plausible,
but require unrealistically large kicks given the 
small amount of mass ejected.}
\label{fig1}
\end{figure}

In short, we generate many random realizations of the binary system, with
different initial masses and different kick velocities. We assume for the kick velocity distribution a prior of equal
probability in logarithmic space. We then evolve the system and check how many
realizations with different initial parameters generate systems which are
consistent with the present observations. Fig. \ref{fig1} depicts the results with different constraints added, namely,
the location within 50pc of the galactic plane, the low orbital eccentricity and  
the small observed proper motion. Finally, the constraint arising from the alignment
of the spin of pulsar A with the orbital angular momentum vector is included: 
Pulsar A's ms spin was already aligned with the orbit before the formation of
pulsar B, having been spun up by accretion, and a significant kick would 
have destroyed this alignment.

We find  that the most likely progenitor mass of pulsar B, $m_B$, was in the 
range $1.47 \div 1.53$ M$_{\odot}$, corresponding to a 
confidence level  of 68\%. The 
associated kick velocity is $\Delta v_{\rm cm} \leq 20$ km/s. Values of the 
progenitor mass up to $\lesssim 1.8$ M$_{\odot}$ lie within the 97\%
probability curve, with associated kick velocities $\lesssim 30$ km/s. 
Finally, a low probability tail extends up to masses
$\lesssim 2.1$ M$_{\odot}$ and velocities $\lesssim
70$ km/s.
The low-mass branch in Fig.\, 1 includes very low progenitor masses, 
$\sim (1.25 \div 1.5)$ M$_{\odot}$, and large kick velocities, $> (30 \div 100)$
km/s. Although kinematically possible, this combination is highly unlikely
as it requires that the ejected mass had very large asymmetry  and we disregard it as spurious.

These values can be understood as follows.
If the  explosion that accompanied the collapse had been spherically
  symmetric, with negligible mass ejection, then the orbital eccentricity
of the system would be simply related to the ejected mass
by  \citep{vdH10}~ $e = {\rm M}_{\rm ej}/({\rm M}_A + {\rm M}_B)$. 
Adopting the gravitational wave-driven evolution of orbital eccentricity  \citep{Eno07} and a lifetime equal to the spindown age of pulsar B, 
the current eccentricity $e \simeq 0.088$ of PSR J0737-3039 can be traced back 
to a maximum $e \lesssim 0.11$ right after the 
formation of pulsar B. By virtue of the above relation, this implies
a maximal mass of $\sim 1.53$ M$_{\odot}$ for the immediate progenitor of 
pulsar B while the minimum, $\sim$ 1.47 M$_{\odot}$, is derived
 using the current orbital eccentricity. 
  
\subsection{Physical mechanism for pulsar B's formation}
\label{subsec:mechanism}

The  progenitor mass of pulsar B is less than 2.1 M$_{\odot}$ at the 99\% confidence level. 
This is less than the minimal progenitor mass required to trigger a 
core-collapse SN \citep{Sm09, Bur13}. It is even below the minimal mass 
of the He core $\sim (2.1 - 2.4$) M$_{\odot}$ needed to ignite O and Ne
burning thus leading to a standard iron-core collapse \citep{Nom87,Dewietal02,
Willems2004}. 
The only remaining plausible scenario is thus that pulsar B
originated from a marginally unstable WD-like degenerate core, which has undergone a   gravitational collapse. 
One possibility for such a collapse is an $e$-capture SN \citep{Myetal80,Nom84}, or the collapse of a hot WD that was initially slightly above the Chandrasekhar limit 
\citep{PiranShavivAstroPh,PiranShavivPRL}. The hot degenerate core might have become unstable because of cooling or, e.g., by accumulating ashes from shell burning at the base of a tenuous envelope (like a stripped down post-AGB star) and then passing the Chandrasekhar limit.
{We will frequently refer to an $e$-capture SN, because this specific case was mostly discussed in the literature. However, the above possibilities stand on equal footing, 
hence the gravitational collapse of an unstable WD-like degenerate core should always 
be understood.}

An $e$-capture SN (or, more generally, a marginally stable WD) is 
expected as the end result of the evolution of stars in a relatively 
narrow range of masses, between 8 and 10 M$_{\odot}$ \citep{Nom84}. 
Evolutionary arguments suggest that $e$-capture SN can preferentially (or only)
occur in binary systems, and involve progenitors with masses up to 
$\sim$ 12 M$_{\odot}$, due to the critical effect that the onset of mass
transfer can have on the evolution of the stellar core \citep{Podsy04, vdH10, 
IbHeg13}. This issue is, however, not yet settled and our arguments do not
depend specifically on this assumption.

Independent of the exact channel through which the unstable WD was formed, its 
collapse released the binding energy of the nascent NS via neutrinos. This 
contributes to the mass difference between the progenitor star and the
resulting NS, thus naturally explaining the relatively small mass (1.25 
M$_{\odot}$) of Pulsar B. The derived mass of the progenitor, (1.47-1.53) M$_\odot$, well exceeds the 
maximum WD mass and a reasonable neutrino mass loss, thus implying that some
mass ejection occurred in the formation of pulsar B, besides the release of the 
binding energy. Precisely how much mass was lost depends on the NS binding energy, which 
is determined by the - yet unknown - NS equation of state (EOS), but it is reasonable to guess that the envelope was 
ejected while the WD collapsed to form the NS. 

\subsection{The progenitor's structure}
\label{sub:progenitorstructure}

\begin{figure}[h!]
\centerline{
\includegraphics[width=10.25 cm]{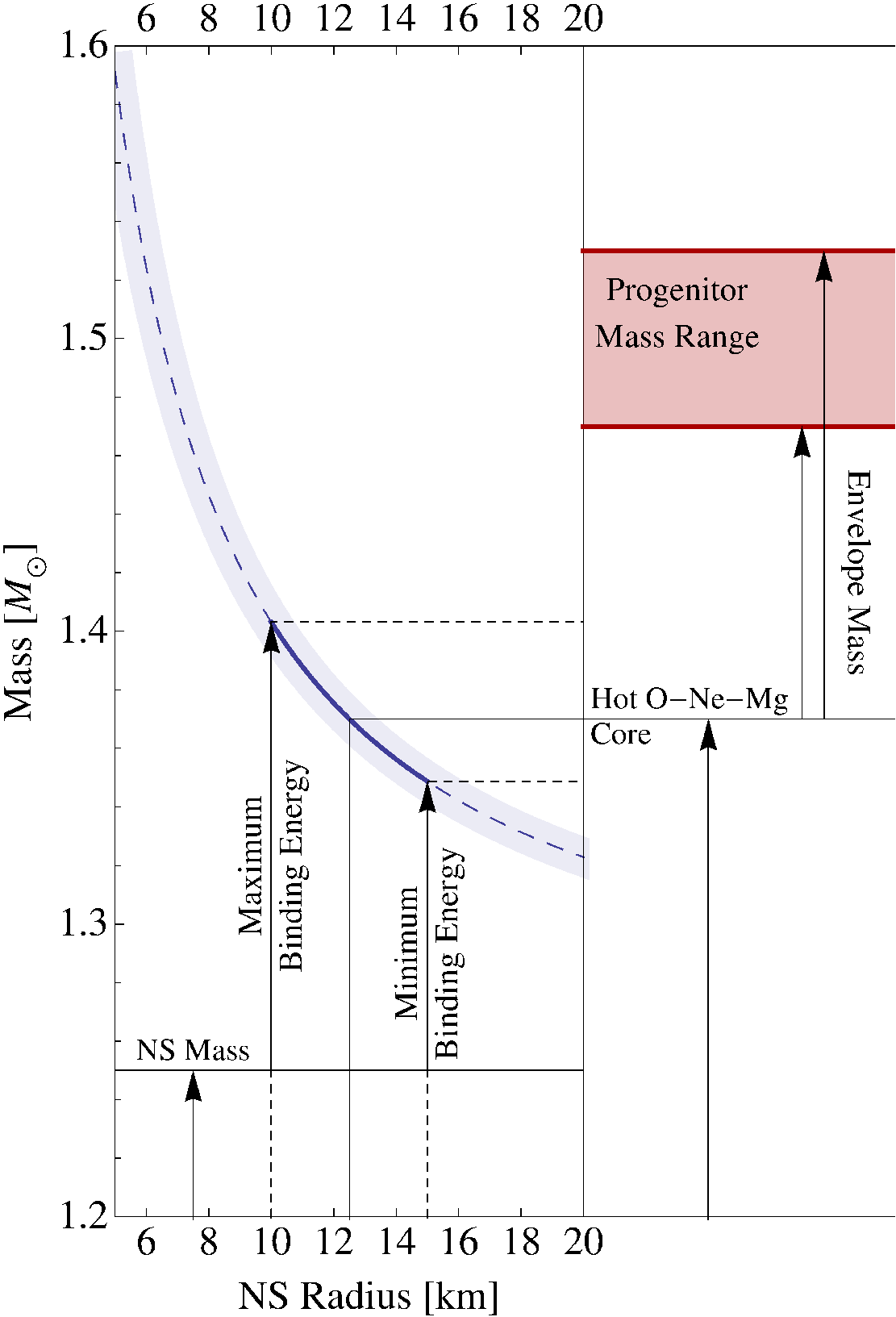}}
\caption{Schematic representation of the structure of Pulsar B's 
progenitor (right-hand side of the figure) and the NS structure (left-hand 
side). The gravitational mass (M$_B$) of Pulsar B is indicated by the 
horizontal line in the left-hand side of the figure. The shaded curve above it 
represents the sum of gravitational mass+binding energy 
(M$_B +{\rm E}_{\rm bind}$) of Pulsar B, where the binding energy is given in 
terms of the NS compactness (thus, its radius) by the expression provided in 
the text. The dashed vertical lines identify the minimum and maximum radius of 
a 1.25 M$_{\odot}$ NS consistent with currently proposed NS  EOSs (cfr. 
\citealt{LattPrak01}). On the right-hand side of the figure the horizontal line
  at 1.37 M$_{\odot}$ indicates our ``chosen'' value for the unstable, 
collapsing WD. The kinematically constrained progenitor mass range is
indicated by the shaded strip above, and the (0.1-0.16) M$_{\odot}$ gap 
indicates the possible envelope mass range at the moment of the collapse. 
Assuming that Pulsar B was formed in the collapse of a hot WD-like core with a mass of 
1.37 M$_{\odot}$, the corresponding binding energy is $\approx 0.12$ 
M$_{\odot}$. This is indicated by the continuous lines on the left-hand side
of the figure, and corresponds to a NS radius of $\approx$ 12 km.}
\label{NS-radius}
\end{figure}

Simple parametric fits to the NS bulk properties can
approximate well  the results of a large class of EOSs 
\citep{LattPrak01, Latt10}. Particularly simple is the relation
between a NS binding energy, $E_{\rm bind}$, and its compactness, $\beta = 
G {\rm M}_B/({\rm R} c^2)$, namely $E_{\rm bind}
\simeq 0.6 {\rm M}_B \beta/(1-0.5 \beta)$. Currently preferred EOSs
indicate NS radii $10~{\rm Km} \lesssim {\rm R} \lesssim$ 12 km
\citep{Ozel12}, corresponding  to  $\beta = (0.125-0.185)$ and 
$0.12 < E_{\rm bind}/{\rm M}_{\odot} < 0.15 $. The resulting 
 ejected mass\footnote{If the age of Pulsar B is comparable to
  its current spindown age, then the mass of its progenitor will be
  close to the maximum allowed value, $m_B \approx 1.53$ M$_{\odot}$, 
and the corresponding ejected mass range becomes M$_{\rm ej} \simeq
(0.13-0.18)$ M$_{\odot}$.} 
M$_{\rm ej} \sim (0.07-0.16)$ M$_{\odot}$.
Note that, considering the whole range of possible EOSs, radii between 10 and
15 Km would be allowed for a 1.25 M$_{\odot}$ NS \citep[see e.g. Fig. 4 in][]{Latt10}.
This would give slightly wider ranges for the physical
parameters, in particular, M$_{\rm ej} \simeq (0.07-0.18) $ M$_{\odot}$.
The limits on the ejected mass, on the NS radius and on its binding energy,
  that can be obtained by our kinematic results, are summarized in Fig. 
\ref{NS-radius}.

The Chandrasekhar mass for a zero-temperature O-Ne-Mg WD is \citep{HamSal61} 
$\approx 1.36 {\rm M}_{\odot}$. This critical mass is slightly modified by finite
temperature effects \citep{Timm96}, leading to a typical value of M$_{\rm cr} 
\gtrsim 1.37$ M$_{\odot}$ at T $\lesssim 10^9$ K.
 The temperature of the WD must have been between $\sim {\rm a~few} \times
  10^8$ K, to allow the growth of the degenerate core through C-burning, and
  $\lesssim 10^9$ K, because further nuclear burnings - that would lead 
to a standard core-collapse - could not be ignited. We thus conclude that 
the immediate progenitor of pulsar B contained a hot, degenerate, O-Ne-Mg 
core with a mass $\sim$ 1.37 M$_{\odot}$ surrounded by a non-degenerate 
envelope of lighter elements, with M$_{\rm env} \sim (0.10-0.16){\rm~M}_{\odot}$, in hydrostatic and thermal equilibrium. This envelope 
is eventually ejected in the explosion while, with our assumed mass, the 
collapsing WD would release a binding energy $E_{\rm bind} =0.12$ M$_{\odot}$,
leading to a final NS mass of 1.25 M$_{\odot}$. This  corresponds to a NS
radius of 12 km, perfectly in line with best  current estimates {(cfr. Fig. \ref{NS-radius}).}

To explore the progenitor's structure we use a simple toy model. 
We approximate the (smooth) transition region between the two zones as a sharp 
interface, located where the gas plus radiation pressure becomes comparable to
(more than half of) the electron degeneracy pressure. This determines uniquely 
the temperature, density and pressure ($T_b, \rho_b {\mbox{~and~}} P_b$) at the 
base of the envelope. The envelope's  structure is then obtained by solving jointly the 
equations of hydrostatic equilibrium and radiative transport, with opacity 
dominated by electron scattering.
In this simple model the outer radius is derived in terms of the inner 
radius (R$_{\rm b}$) and the pressure scale height at the base of the
envelope, $\phi = 4 P_{\rm b}/(\rho_{\rm b} g_{\rm b} {\rm R}_{\rm b})$, such 
that ${\rm R}_{\rm out} = {\rm R}_{\rm  b}/(1- \phi)$. The total envelope mass 
is also determined in terms of these quantities, 

\begin{equation}
{\rm M}_{\rm env} = 
4 \pi \rho_{\rm b} (R_{\rm b}/\phi)^3 \left[- {\rm ln}(1-\phi) - \phi - 
(\phi^2 /2) - (\phi^3/3)\right] \, . 
\end{equation}

\begin{figure}[h!]
\centerline{\includegraphics[width=9cm]{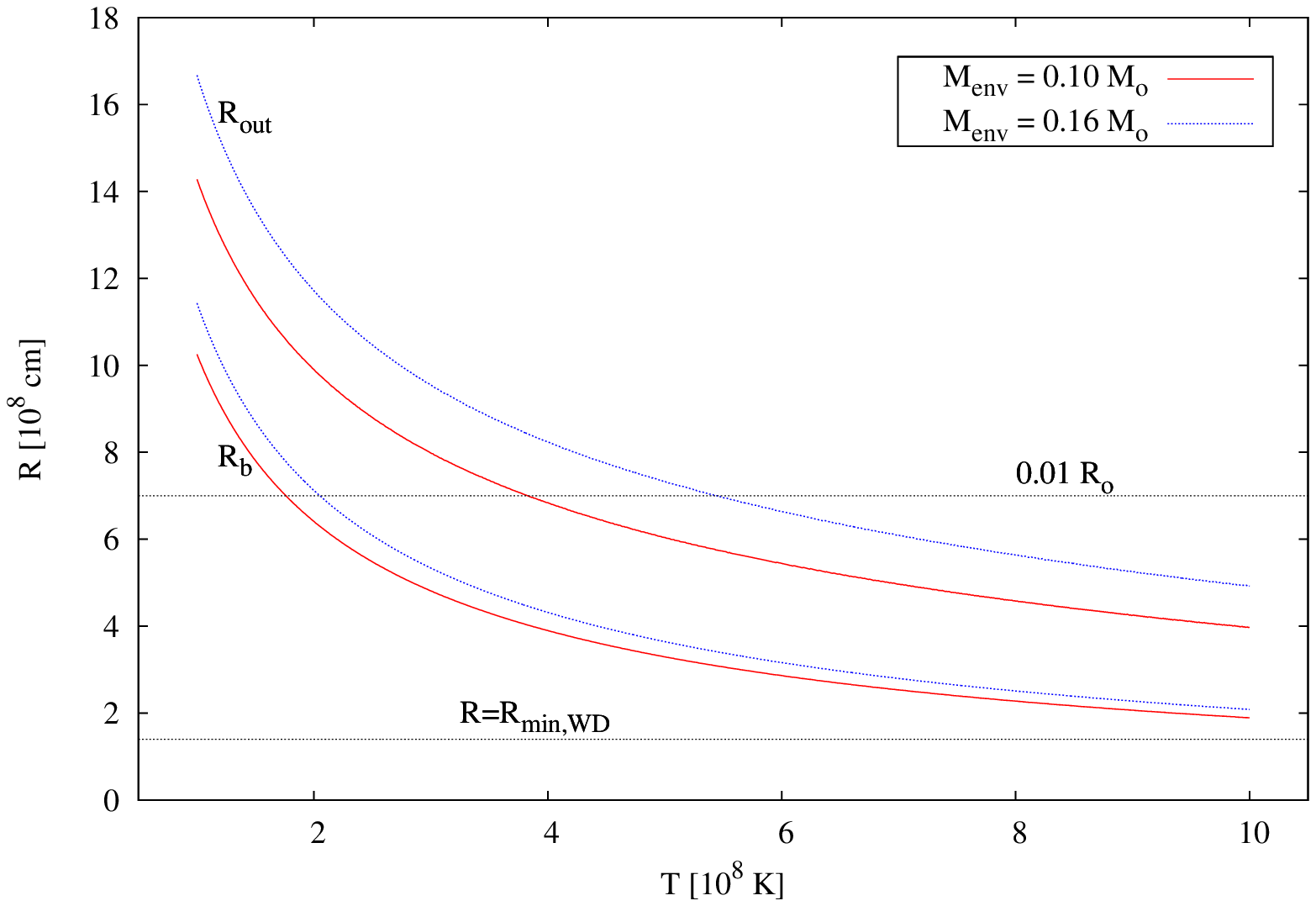}
  \includegraphics[width=9cm]{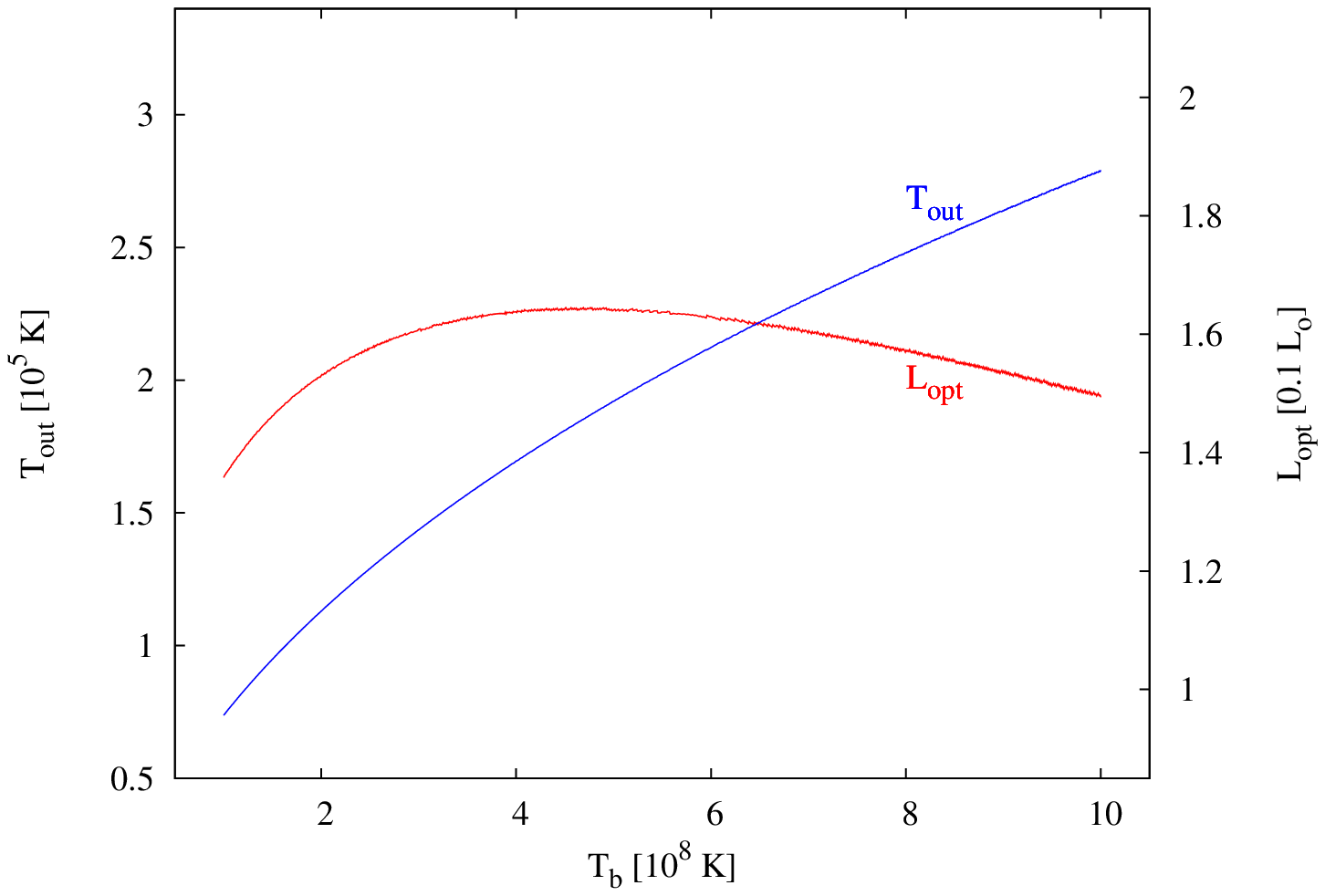}} 
\caption{{\textit{Left Panel:}} The core radius (R$_{\rm b}$) and the 
envelope outer radius (R$_{\rm out}$) as a function of the temperature at the 
core/envelope border (T$_{\rm b}$) for an hot, degenerate core surrounded by a 
non-degenerate envelope. Two curves are shown for both radii, corresponding to
the minimum and maximum value of the envelope mass, as derived in the
text. {\textit{Right Panel:}} The temperature at the surface of the
envelope (T$_{\rm out}$) and the corresponding luminosity in the optical band
(4000-7000) \AA, in units of 0.1 L$_{\odot}$, for the case with M$_{\rm env}
= 0.1$ M$_{\odot}$. The temperature scale is reported on the left $y$-axis, 
while the luminosity scale is reported on the right $y$-axis.}
\label{fig0}
\end{figure}

Fixing the total envelope mass, we can thus 
determine analytically the required values of ${\rm
  R}_{\rm b}$ and $\phi$, hence ${\rm R}_{\rm out}$, as a function of 
${\rm T}_{\rm b}$. Fig. \ref{fig0} depicts the inner
and outer radii of the envelope as a function of T$_b$, for two 
different values of the envelope mass, corresponding to the minimal and
maximal values of the progenitor mass derived from the kinematic
constraints (1.47 and 1.53 M$_\odot$ respectively). At smaller temperatures R$_{\rm b}$
seems implausibly large for a critical-mass WD; for $T \geq 5 \times 10^8$ K, the 
core radius is less than twice the radius  of a zero-temperature WD ($\sim 1.4 \times 10^8$ cm), roughly consistent with the (non-negligible) effects on radius due to partial
degeneracy at such high T. The corresponding values of ${\rm R}_{\rm out}$ 
are $\sim$ twice as large as the inner radius, with a typical size $\sim 
0.01$ R$_{\odot}$.

Finally, the surface luminosity of the WD is determined as a function of 
${\rm T}_{\rm b}$ as ${\rm L}_{\rm WD} = {\rm L}_{\rm Edd} 
(P_{\rm b, rad} /P_{\rm b})$, where L$_{\rm Edd} = 4 \pi G M c/ \kappa$ is the 
Eddington luminosity, (${\rm P}_{\rm b, rad}$) the radiation pressure and 
(${\rm P}_{\rm b}$) the total pressure at the base of the envelope. 
The implied surface temperature and optical luminosity as functions of 
${\rm T}_{\rm b}$ are shown, in the right panel of Fig. \ref{fig0}, for
the case ${\rm M}_{\rm env} =0.10~{\rm  M}_{\odot}$. The system was
 rather dim in the optical band, with a typical luminosity peak of $\sim 0.15$
 L$_{\odot}$. Due to its high temperature it was, however, much brighter than 
a typical WD in the UV, where most of the emission occurred, with an estimated 
luminosity $\sim 10^{36}$ erg s$^{-1}$.

{Our approximation doesn't include important effects, such as neutrino cooling and heat conduction in the outer layers. Furthermore, we treat the transition region in a crude way. More accurate results would require numeircal modeling of these effects. However, we don't expect these effects to change qualitatively our conclusions.}

\section{Kinematic considerations: the first SN and pulsar A progenitor's
  mass }
 \label{pulsarA}

Remarkably, the slow proper motion of the system enables us also  to put an
  interesting limit on $m_A$, the mass of the stellar progenitor of pulsar A just before it collapsed. 
  Let $\tilde{m}_B$ be the mass of the star B when the first SN took 
place\footnote{This is different from, {\textit{i.e.}} larger than $m_B$, the mass of
    star B right before it collapsed.}.
  Mass ejection  in this first SN results in a  recoil
  velocity of the center-of-mass  \citep{PiranShavivAstroPh}:
\begin{equation}
\label{eq:centerofmass}
\Delta {\mathbf{v}}_{\rm cm} = \frac{{\rm M}_{\rm tot,i}-{\rm M}_{\rm tot,f}}{{\rm M}_{\rm tot,f}}~
\frac{\tilde m_B}{{\rm M}_{\rm tot,i}}~ {\mathbf{v}}_{\rm rel,i} , 
\end{equation}
where M$_{\rm tot,i}\equiv \tilde m_B+m_A$ and M$_{\rm tot,f}\equiv \tilde m_B + M_A $,  represent the total mass in the
    binary \textit{before} and
    \textit{after} the first SN, respectively. 
$v^2_{\rm rel,i} = G {\rm M}_{\rm tot,i}/a_i$ is
    the relative (orbital) velocity prior to the SN (assuming a circular 
orbit) and $a_i$ is the corresponding orbital separation.

The observational constraint
$\mid \Delta v_{\rm cm} \sin
\theta_{\rm los} \mid \leq \mid v_{{\rm obs},\perp} \mid$, where $\theta_{\rm los}$ is
  the angle between the direction of ${\mathbf{v_{{\rm rel,i}}}}$ and our
  line-of-sight, limit the amount of mass that could have been
  ejected in the first SN. This in turn puts an upper limit  on
the total mass of the binary {\textit{before}} the first SN went off, 
${\rm M}^{({\rm max})}_{\rm tot,i}$, which is a function of the parameters ($\tilde m_B,
\sin \theta_{\rm los}; a_i, v_{{\rm obs},\perp}$).
This upper limit also constrains the mass ratio in the binary at the
  moment of the first SN. To see this let's impose the condition $m^{({\rm
      max})}_A > \tilde m_B$,  which of course implies ${\rm M}^{({\rm
      max})}_{\rm  tot,i} > $ 2 $\tilde m_B$.
We obtain:
\begin{equation}
\label{eq:sintheta}
\sin^2 \theta_{\rm los} < 0.224~\frac{(a_i/{\rm AU})}{(\tilde m_B/M_\odot)}
~\left(\frac{v_{{\rm obs},\perp}}{10 ~ {\rm km/s}}\right)^2 \left[ \frac{\tilde m_B+ {\rm M}_A}{\tilde m_B - {\rm M}_A}\right]^2 \, ,
\end{equation}
which  holds only if the center-of-mass of the system obtained a recoil
  velocity that was directed very close to our line of sight.
When the angle $\theta_{\rm los}$ is drawn from a random distribution, the
    probability that $\theta_{\rm los} \leq \overline{\theta}_{\rm max}$ is simply
    P$(\theta_{\rm los} \leq \overline{\theta}_{\rm max}) = 1- \cos \overline{\theta}_{\rm max}$.
    Combining this with eq. \ref{eq:sintheta}, the conditional probability 
P(M$_{\rm tot,i} > 2 \tilde m_B \mid \tilde m_B) $ can be derived, for fixed
values of the parameters ($v_{{\rm obs},\perp}$, $a_i$, M$_A$). Finally, 
assuming a flat prior for the values of $\tilde m_B$ between 8 
and 12 solar masses,
the conditional probability can be marginalized over $\tilde{m}_B$ to derive that the total probability
P$({\rm M}^{({\rm max})}_{\rm tot,i} > 2 \tilde m_B) \simeq 0.02$. 
The condition $m_A > \tilde{m}_{\rm B}$ that 
  requires that the center-of-mass recoil velocity was pointed less than 
$\approx 12^{\circ}$
  away from our line of sight ($\simeq 2$ \% probability) is  highly fine-tuned.
We conclude  that when the progenitor star of pulsar A exploded, it must have 
been less massive than its companion.

This  conclusion implies that a phase of significant 
mass transfer must have occurred  prior to the first SN. 
This phase reversed the original mass ratio, at the same time bringing the progenitor of pulsar B to
    the right mass range for producing the second NS. The low proper motion of
     PSR J0737 thus provides evidence of this early binary evolutionary
     phase, based on the dynamical properties of its late descendants.

\section{Mass transfer in later evolutionary phases}
\label{masstrasfer}

Based on the measurement of a very low proper motion of PSR J0737-3039, on its
low galactic latitude and the near alignement of pulsar A's spin  with the
normal to the orbital plane, we concluded that pulsar B was formed in the
collapse of an unstable WD-like degenerate core, thus providing evidence for a new formation
channel for NSs. 
We also argued that the slow proper motion of the system represents by itself
a (first) dynamical evidence for a phase of mass 
transfer between the stellar progenitors, prior to the formation of the first NS in the system. 
The dynamic properties of the double pulsar also give interesting indications
about later stages in the evolution of the progenitor system.

\subsection{Common Envelope and Roche Lobe Overflow}
\label{CE-RLOF}
With a total mass of $\sim$ 2.6 M$_{\odot}$ and a period of $\sim$ 2.45 
hrs, the current orbital separation of the two NSs, $a \lesssim 9 \times 
10^{10}$ cm, is only slightly larger than the solar radius. As such it could not accomodate the
  progenitor stars of the two pulsars, that must have been both much larger
  than our sun. Significant orbital shrinkage must  have occurred  before
the formation of 
pulsar B, implying that the binary went through a
  phase of common-envelope (CE) after the formation of pulsar A.
The CE phase stripped the 
envelope of the $\sim (8-12) $ M$_{\odot}$ star and
eventually left pulsar A orbiting around an $\sim (3.5-6)$ M$_{\odot}$ He star 
in a close binary. 
At the end of core He burning, the He star left the main sequence and a further 
stage of Roche-lobe overflow (RLOF) occurred in this system (cf. 
\citealt{Dewi2003, Dewi2004}). Depending 
on the mass of the He star, this latter stage is expected to last $\sim 10^3
\div 10^5$ yrs, lower mass stars corresponding to a longer (nuclear)
evolutionary timescale  \citep{Dewietal02, Dewi2004}. When 
mass transfer ends most of the He envelope is lost, leaving behind the
degenerate O-Ne-Mg core surrounded by a tenous layer of material.

\subsection{Acceleration of the millisecond pulsar}

The fast spin (23 ms) of pulsar A and its alignment
with the normal to the orbital plane are a natural consequence of the final 
phase of RLOF accretion in the above scenario. Mass flowing through the inner Lagrangian point will 
settle into a disk and feed angular momentum to the NS at the
 rate $\dot{{\rm M}} ({\rm G} {\rm M}_A {\rm R}_a)^{1/2}$. Here $\dot{{\rm M}}$
  is the mass accretion rate, M$_A$ is the mass of Pulsar A and R$_a$ the magnetospheric 
(or Alfv\'{e}n) radius, where the Keplerian flow breaks down and 
matter co-rotates with the magnetosphere. The latter can be written 
as  \citep{Pat12}~R$_a = 26.8 ~{\rm km} 
~\xi ~(B_9 R^3_6)^{4/7} (\dot{M}/\dot{M}_{\rm E,He})^{-2/7} ({\rm M}_A/M_{\odot})^{-1/7}$, where 
  $\xi \sim (0.3 \div 1)$ parametrizes the relative thickness of the
transition 
layer between Keplerian flow and corotation  \citep{PsaCha99}, $\dot{{\rm
      M}}_{\rm E,He} \approx 3 \times 10^{-8}$ M$_{\odot}$ yr$^{-1}$ is 
 the Eddington limit for He and $Q_x
    \equiv (Q/10^x)$ [c.g.s. units] for any quantity $Q$. The current 
angular momentum of pulsar A is $\approx 2\times 10^{47}$ g cm$^2$ s$^{-1}$
(m$_A/$ M$_{\odot}$) R$^2_{A~6}$, 
where the NS moment of inertia $I = k^2_A {\rm M}_A R^2$ and we took $k^2_A =1/3$. This implies that    in order to spin up the NS to its current 23 ms period, the accretion phase must  have lasted $\tau_{\rm acc} \simeq (1.2 \div 2.1)  \times 10^5$ yrs, during which $\sim (3.6 \div 6.3) \times 10^{-3}$    M$_{\odot}$ were accreted. This small amount of mass is consistent with the rather low mass of pulsar A.

The value of $\tau_{\rm acc}$ derived here is somewhat longer than previous
estimates for He stars that are in the (4.5$\div 6$) M$_{\odot}$  range
\citep{Dewietal02, Dewi2004}. Those estimates, however, assumed that 
pulsar B was formed in a core collapse SN, hence the degenerate core of the He 
star had to exceed $\sim 2.1$ M$_{\odot}$. Our result is qualitatively more 
in line with the progenitor He star having been less massive, tentatively 
$\sim (3 \div 4)$ M$_{\odot}$, thus developing a lighter core that could not 
trigger a core-collapse SN. 

The weak magnetic field of pulsar A ($\sim 7 \times 10^{9}$ G) also
  fits the picture of accretion-induced decay of NS exterior
  fields. Empirically, the field evolution is found to scale with the
  accreted mass as  $B(t) = B_0/(1+\Delta M/k_m)$, with the 
observationally determined constant  $k_m \approx 1.25 
\times 10^{-5}$ M$_{\odot}$ \citep{Dewi2004, FraWijBro02}. In this picture, 
pulsar A's field would have decayed by a factor $\sim
  (300 \div 500)$, making its original value consistent with those typical of NSs. 

\section{The spin of pulsar B}
\label{spinB}

Another remarkable property of PSR J0737-3039 is that, while pulsar A's spin 
is aligned with the orbit, pulsar B's spin appears to be misaligned at 
130$^\circ$. We turn now to identify the mechanism that produced this misalignment by 
considering the possible evolutionary steps of  Pulsar B's progenitor  up to the final collapse.
A critical process that controls the spin of the NS is the earlier 
coupling between the stellar core and the envelope. This coupling 
determines the spin of the WD just before its collapse and hence it influences the NS
spin. 
We thus consider this issue before addressing the 
question of the origin of pulsar B's misaligned spin.

As discussed in $\S$ \ref{CE-RLOF}, Pulsar B's progenitor experienced
 at least two phases of intense mass loss during which a large amount of
angular momentum was  removed from its envelope. If the core were
effectively coupled it would have also been affected by the external torque, loosing
its angular momentum. 
The tidal field of Pulsar A would also tend to synchronize (and align)
  the rotation of the envelope of star B to the orbit, particularly during
  RLOF \citep{Zahn77}.
If the (radiative) core were effectively coupled to the envelope, then a WD 
with a spin period of the order of the orbital period would result.
A decoupled core, on the other hand, would retain its 
original angular momentum and the final orientation (and magnitude) of the WD 
spin would only depend on initial conditions.

The low degree of differential rotation found in the core of the sun and 
of some massive stars (\citealt{PhiSpru98} and references therein), provides 
observational indications for a rather generic coupling between core and 
envelope \citep{MaedMey12, ChaTal05}. 
The slow measured spins of isolated WDs give independent observational support 
for the occurrence of strong losses of angular momentum in the cores of their 
stellar progenitors  \citep{Kaw04, Kaw05, Lan07, Suj08, ChaFoBra09}. These 
findings have in turn stimulated  theoretical investigations on the 
possible mechanisms that enforce efficient coupling between core and envelope in
massive stars  \citep{Zahn92, TalCha05}.
One specific mechanism that received particular attention invokes the 
effect of internal magnetic stresses  \citep{PhiSpru98}, as these are generally
amplified by significant differential rotation and efficiently oppose its
growth. Convective motions are expected to develop in the transition region 
between the (radiative) core and the envelope of the He star
\citep{Dewietal02,Ivan03},  and these might also play a significant role in
the synchronization of the core \citep{Zahn96}. 
Overall  it seems that there is observational evidence and theoretical reasoning to expect that the core and the envelope are coupled. Still as 
this is not certain \citep{Cors11, Becketal12}, we explore in the following 
both possibilities.

Returning to the  spin misalignment of pulsar B we note that it can be explained in one of two possible ways:
 \textit{1})  Pulsar B progenitor's spin was  misaligned relative to the
 orbital plane before the collapse, and the spin didn't change when the NS was 
formed;  \textit{2}) Pulsar B progenitor's spin was aligned with the orbital 
plane, and the spin direction was changed during the NS formation process. 
We turn now to consider each one of these scenarios.

 {\bf Scenario 1:}  In this scenario pulsar B progenitor's spin was  misaligned relative to the
orbital plane before the collapse and the current spin direction of pulsar B was inherited from
its  progenitor star.  The final collapse  of the unstable WD did not affect it.
It is reasonable to assume that, in the original binary star system, 
the two spins were aligned with the orbital angular momentum, 
$\mathbf{{\rm {\bf L}}_{\rm orb}}$. Depending on the coupling between the core and the envelope, there are 
two moments in the  subsequent evolution of the system when the spin vector of B could have 
been tilted and we consider the two cases seperately.

{\bf Strong coupling between the stellar core and the envelope}: The coupling
implies that the spin of the core was aligned with the stellar spin, 
which in turn, was aligned with the angular momentum of the orbit. In this 
case the tilt could have resulted during the ejection of  the envelope 
accompanying the formation of the WD. 
It was suggested that this process could 
be responsible for the observed distribution of natal spin periods and kick 
velocities of isolated WDs, respectively 
$\sim (0.15- {\rm ~a~few})$ days and $\sim (1-{\rm~a~few})$ km/s 
\citep{Spruit98, Davis06, Heyl07}. 

In isolated WDs the envelope is ejected in a $\sim 10^4$ yr-long super-wind 
phase that sets in when the progenitor star reaches the AGB  phase \citep{Spruit98}.
However, the onset of mass transfer onto Pulsar A might have avoided
  precisely this phase in the progenitor of Pulsar B, changing the fate of its
  degenerate core \citep{Podsy04, vdH10}. 
In fact, the unstable WD was formed in the core of the He star while its 
envelope was ejected via RLOF, and the ensuing mass transfer phase. It is this 
different physical situation that we address now.

The ejected mass flows through the inner Lagrangian point towards the
accreting neutron star, mostly in the radial direction. While a small degree of 
anisotropy of the flow would possibly result in a series of random kicks, like
in the case of AGB super-winds  \citep{Spruit98}, two major differences with the 
latter case must be stressed. On the one hand, RLOF is expected to be much 
smoother than the highly variable and inhomogeneous winds of AGB stars, thus 
leading to smaller individual kicks (if any). On the other hand, the Roche 
radius limits the size of the mass donor to be $\lesssim $R $_{\odot}$, much 
less than the radius of a typical AGB star ($> 10^2$ R$_{\odot}$). This
strongly reduces the ``lever arm'' of the kicks, hence their effectiveness.
Adopting the same scaling law as for AGB winds, the total angular momentum 
deposited onto the WD at the end of the mass transfer by fall back of the residual 
envelope material 
is given by  \citep{Spruit98}:
\begin{eqnarray}
\label{eq:deposited-L}
{\rm L}_{\rm F} & \sim &\frac{\epsilon}{\sqrt{N}} ~{\rm R}_* c_s {\rm
  M}^{1/2}_{\rm env,0} {\rm M}^{1/2}_{\rm env,F} \sqrt{3~k^2_*/2} \nonumber \\
  & \approx & 3 \times 10^{43} ~\frac{{\rm g~cm}^2}{{\rm s}}~\epsilon_{-2} 
\left[\left(\frac{{\rm M}_{\rm env,0}}{2~{\rm M}_{\odot}}\right) \left(\frac{{\rm
      M}_{\rm env,F}}{0.1~{\rm M}_{\odot}}\right) \left(\frac{{\rm
      R}_{\rm RL}}{{\rm R}_{\odot}}\right)^3 \left(\frac{{10^5~\rm yrs}}{\tau_{\rm acc}}\right)\right]^{1/2}~{\rm T}^{1/4}_{\rm s,4} \, .
\end{eqnarray}
In the above M$_{\rm env,0}$ is the envelope mass at the beginning of mass
transfer, M$_{\rm env,F}$ its residual mass at the end of mass transfer -
normalized to the envelope mass estimated in
sec. \ref{sub:progenitorstructure} 
- and T$_{\rm surf}$ is the surface temperature of the expanded star.
The degree of anisotropy of the flow, $\epsilon_{-2}$, was normalized to 
$10^{-2}$ and the velocity of the ejected material was assumed to be the local 
sound speed at the stellar surface.
This  estimate shows that this mechanism  fails to provide to 
the WD enough {\textit{misaligned}} angular momentum to  account for 
pulsar B's current spin. As such, this explanation should also be ruled out.

{\bf{No coupling between the stellar core and the envelope:}} 
If the core of pulsar B's progenitor was never coupled to 
the envelope it would have maintained its original spin direction.
If the orbital plane  changed abruptly by the kick imparted when the first SN exploded,
then after the collapse, the spin of pulsar B would eventually reveal the original
orientation of the orbital plane.  Accretion onto pulsar A would, on the other
hand, lead to  the alignment of its spin with the new orbital plane.

The angular momentum that is transferred by the natal kick to the orbit is 
$\Delta {\rm {\bf{L}}_{\rm kick}} = {\rm M}_A~\mathbf{v}_{\rm  kick}
\times \mathbf{a_1}$, where $a_1 = a~ \tilde m_B/{\rm
  M}_{\rm tot,f}$ is the distance of the newly formed NS (Pulsar A) 
from the center of mass.  To affect the orientation of the orbital plane, this should exceed the 
original angular momentum, ${\rm {\bf{L}}}_{\rm orb} = (m_A \tilde m_B/{\rm
  M}_{\rm tot,i})~ \mathbf{a_i} \times \mathbf{v}_{\rm orb,i}$. The condition 
$\Delta \mathbf{{\rm L}_{\rm kick}} > 
\mathbf{{\rm L}_{\rm orb}}$ ultimately requires that
\begin{equation}
\label{eq:Lkick-condition}
a_i > \left(\frac{m_A}{{\rm M}_A}\right)^2
  \left(\frac{{\rm M}_{\rm tot,f}}{{\rm M}_{\rm tot,i}}\right) \frac{G
    {\rm M}_{\rm tot,f}}{v^2_{\rm kick}} \approx 4~{\rm AU}
  \left(\frac{m_A}{6~{\rm M}_{\odot}}\right)^2 \left(\frac{{\rm
      M}_{\rm tot,f}}{9.35~{\rm M}_{\odot}}\right) \left(\frac{v_{\rm kick}}{200~{\rm km/s}}\right)^{-2} \left(\frac{{\rm
      M}_{\rm tot,f}}{{\rm M}_{\rm tot,i}}\right)  .
\end{equation}
This is $\gtrsim 800$ R$_{\odot}$ and it exceeds the maximal radius to which
    an M $< 12$ M$_{\odot}$ star is expected to expand on the giant branch. 
Since the system went through a CE phase, pulsar A must have been 
within the companion's envelope at some stage of the evolution. Hence, 
constraint (\ref{eq:Lkick-condition}) argues against the viability of this 
scenario.

This conclusion is further strengthened if we account for 
the actual angle between the spin of pulsar B and the normal to the orbital plane,  
$\pi/2+\beta \approx 130^{\circ}$. In this scenario this corresponds to the angle by which $\mathbf{{\rm   L}_{\rm o}}$ was tilted by the kick. In the sum {\bf{\rm
    L}}$_{\rm F} =$ {\bf{\rm L}}$_{\rm o} + \Delta {\bf{\rm L}}_{\rm kick}$,
the  vector $\Delta {\bf{\rm L}}_{\rm kick}$ must  be tilted away from {\bf{\rm
    L}}$_{\rm o}$ by an angle  $\gamma > \beta$, which implies that only
$\lesssim $ 15\% of the possible random orientations of $\Delta{\rm
  \bf{L}}_{\rm kick}$ are allowed.
After some manipulation, we derive the following condition on the modulus
of $\Delta {\rm L}_{\rm kick}$
\begin{equation}
\label{eq:vector-L}
\mid \Delta {\rm L}_{\rm kick} \mid =  \mid {\rm L}_{\rm o} \mid
\left(\frac{1}{1- \sin 2\gamma}\right)^{1/2} =
\mid {\rm L}_{\rm o} \mid \hat{g}(\gamma) 
\end{equation}
where $\beta \approx 40^{\circ}$ was used and  $40^{\circ} < \gamma <
90^{\circ}$. In eq. (\ref{eq:Lkick-condition}), the rhs should thus
    be multiplied by the coefficient $\left[\hat{g}(\gamma)\right]^2 >1$.
In particular, $\hat{g}(\gamma)^2 >2$  for $ 40^{\circ} < \gamma \lesssim
80^{\circ}$,
making the required orbital separation impossibily large. This range of
$\gamma$ leaves out only $\approx$ 5\% of the orientations allowed by 
$\gamma > \beta$, and less than $\approx$ 1\% of all the possible random orientations 
of $\Delta{\rm \bf{L}}_{\rm kick}$. We conclude that this scenario is ruled out: 
a natal kick to pulsar A would not have been able to tilt the orbital plane by the
required amount.

We conclude that  no known mechanism could have
tilted the spin of the progenitor star before the formation of the WD, whether
its core was coupled or not to the envelope. Therefore the current spin  direction of pulsar B must have 
been set during the collapse. This is scenario 2). 

{\bf Scenario 2: The spin direction was changed during the NS formation 
process.}
In order for the spin direction to change significantly, the collapse 
must have imparted to the nascent NS angular momentum in excess of that 
available to the progenitor. The current spin of pulsar B provides a lower 
limit of $\simeq 2 \times 10^{45}$ g cm$^2$ s$^{-1}$ to the required angular 
momentum. 
However,  the low proper motion of the binary also puts a robust
 upper limit on the angular momentum that could have been imparted to the NS.

An arbitrarily oriented, off-center kick could impart spin angular
  momentum to the nascent NS. Given that a linear momentum M$_{\rm tot}
  \Delta v_{\rm cm}$ would also be transferred to the binary, with M$_{\rm
  tot}= M_A + M_B$, the maximum spin that can be given to pulsar B in this way is
$\Delta L \approx R_{\rm kick} \Delta v_{\rm cm} M_{\rm tot}$, where $R_{\rm kick}$
  is the distance from pulsar B's rotation axis at which the kick is imposed.  
Adopting the current upper limit on $\Delta v_{\rm cm}$ we  obtain:
\begin{equation}
\label{eq:lkick_N1}
\Delta {\rm L} \approx 
5.2 \times 10^{45} ~{\rm ~g~cm/s}~ \left(\frac{{\rm R}_{\rm kick}}{{\rm
    R}_B}\right) \frac{\Delta v_{\rm cm}}{10~{\rm km/s}} \left(\frac{{\rm R}_B}{10~{\rm km}}\right) \left(\frac{{\rm M}_{\rm tot}}{2.6~{\rm
    M}_{\odot}}\right) \, .
\end{equation}

This modest amount of angular momentum  is sufficient to account for
 pulsar B's tilted spin \textit{if the progenitor WD was slowly rotating}, and if 
 R$_{\rm kick} \gtrsim (1/2)$ R$_B$ $(\Delta v_{\rm  cm}/10~{\rm km/s})^{-1}$.  
This shows that the putative kick would have to be significantly off-centered to 
match the current rotation of Pulsar B. Actually,  the maximal angular
momentum allowed by eq. (\ref{eq:lkick_N1}) exceeds the current angular
momentum of Pulsar B only by a factor of a few\footnote{Under the
reasonable assumption that R$_{\rm kick}$ does not exceed R$_{\rm NS}$ by a large factor.}.

Assuming that the maximum angular momentum given by the kick is roughly
  twice\footnote{See below.} the value  obtained in eq. 
(\ref{eq:lkick_N1}), we can derive an upper limit to the allowed spin rate of 
the WD progenitor. 
Upon writing $ {\rm L}_{\rm WD} = I_{\rm WD} \Omega_{\rm WD} $ we obtain
\begin{equation} 
\label{eq:fastest-possible}
\Omega_{\rm WD} < 4.5 \times 10^{-4}~{\rm rad/s}
\left(\frac{\Delta v_{\rm cm}}{10{\rm km/s}}\right) \left(\frac{{\rm
    R}_B}{10{\rm km}}\right)
\left(\frac{{\rm M}_{\rm  WD}}{{\rm M}_{\rm Ch}}\right)^{-1} 
\left(\frac{{\rm R}_{\rm   WD}}{2\times 10^8{\rm cm}}\right)^{-2} 
\left(\frac{{\rm M}_{\rm tot}}{2.6{\rm M}_{\odot}}\right)
\end{equation}
which corresponds to a minimum spin period of  $\sim$ 3.1
hrs $\sim 0.13$ d. 
The fact that this limit is reminiscent of the orbital period of the
system prior to the second SN, suggests that tidal synchronization and
  alignment of star B's spin might have played an essential role in
setting the final magnitude and orientation of the angular momentum of 
the unstable WD\footnote{The limit given by Eq. \ref{eq:fastest-possible}
also well matches the fast-rotation end in the
  distribution of spin periods of isolated WDs \citep{Heyl07}. This
  distribution is thought to be caused by a combination of strong mass
  losses and efficient core-envelope coupling in AGB stars \cite{Spruit98}. 
This condition might well have occurred in the evolution of Pulsar B's
progenitor despite the very different environment.}.

An off-center kick could be part of
the collapse itself if the outgoing neutrino flux in the early convective
phase of the proto-NS had some small, yet finite, degree of anisotropy
\citep{Keil96,PhiSpru98,Bur13}.
The kick due to neutrino anisotropies would naturally be applied at the
neutrinosphere, at a radius R$_{\nu} = f_{\nu} {\rm R}_{\rm NS}$, with $f_{\nu} \sim$ a few.
The resulting angular momentum
would be enough to account for pulsar B's current  spin if $\Delta v_{\rm cm} 
\gtrsim 2~(f_{\nu}/2)^{-1}$ km/s (eq. \ref{eq:lkick_N1}). 
It is interesting to note that these numbers well 
match the kick estimates based on the current picture of neutrino-driven 
proto-NS
convection \citep{JaMu94,PhiSpru98,Jan12,Bur13}, in which the neutrino
anisotropy is only $\sim$ a few \% on the scale of convective cells. This
further supports the overall consistency of this scenario.

We conclude that the tilted spin of Pulsar B can be well explained
  by an off-centered kick imparted to the proto-NS during the collapse, most likely due
  to a small degree of anisotropy in the outgoing neutrino flux. Independent
  of the actual cause of the kick, this scenario has
  two major implications: \textit{i)} the NS was born with a relatively slow
  rotation, with an initial spin period $\gtrsim  0.5$ s corresponding to $f_{\nu}=2$
    in eq. (\ref{eq:lkick_N1}). The NS received a mild kick which 
changed its original spin direction and imparted a slow proper motion to the 
binary;
  \textit{ii)} The progenitor WD retained only a small amount of its original angular
  momentum, such that the mild kick could affect the spin of
  the nascent NS. This supports the occurrence of
 a phase of effective core-envelope coupling in the progenitor (He) star.

\section*{Conclusions}
\label{conclusions}
We have considered the peculiar dynamic properties of the binary pulsar
  PSR J0737-3039. These dynamic properties pose strong constraints on the
  evolutionary path that has led to the formation of this system and lead to a well 
  determined evolution. Unlike the ``first" double pulsar PSR 1913+16, whose observed 
  masses and orbital parameters can be explained as arising from standard core collapses of two massive 
  stars \citep{BurrowsWoosley86}, the dynamical parameters here rule out this ``standard scenario" and pose
  very stringent constraints leading to a unique evolutionary path. 
  A summary of the main evolutionary steps followed by this binary, since
when it hosted massive stars up to when Pulsar B was formed, is depicted in 
Fig. \ref{fig:evolution}. Also indicated in this figure are the corresponding 
changes in the orientation of the stellar spins (and of the orbital plane).
We now summarize the main findings:

\begin{figure}[t!]
\centerline{\includegraphics[width=10.35cm]{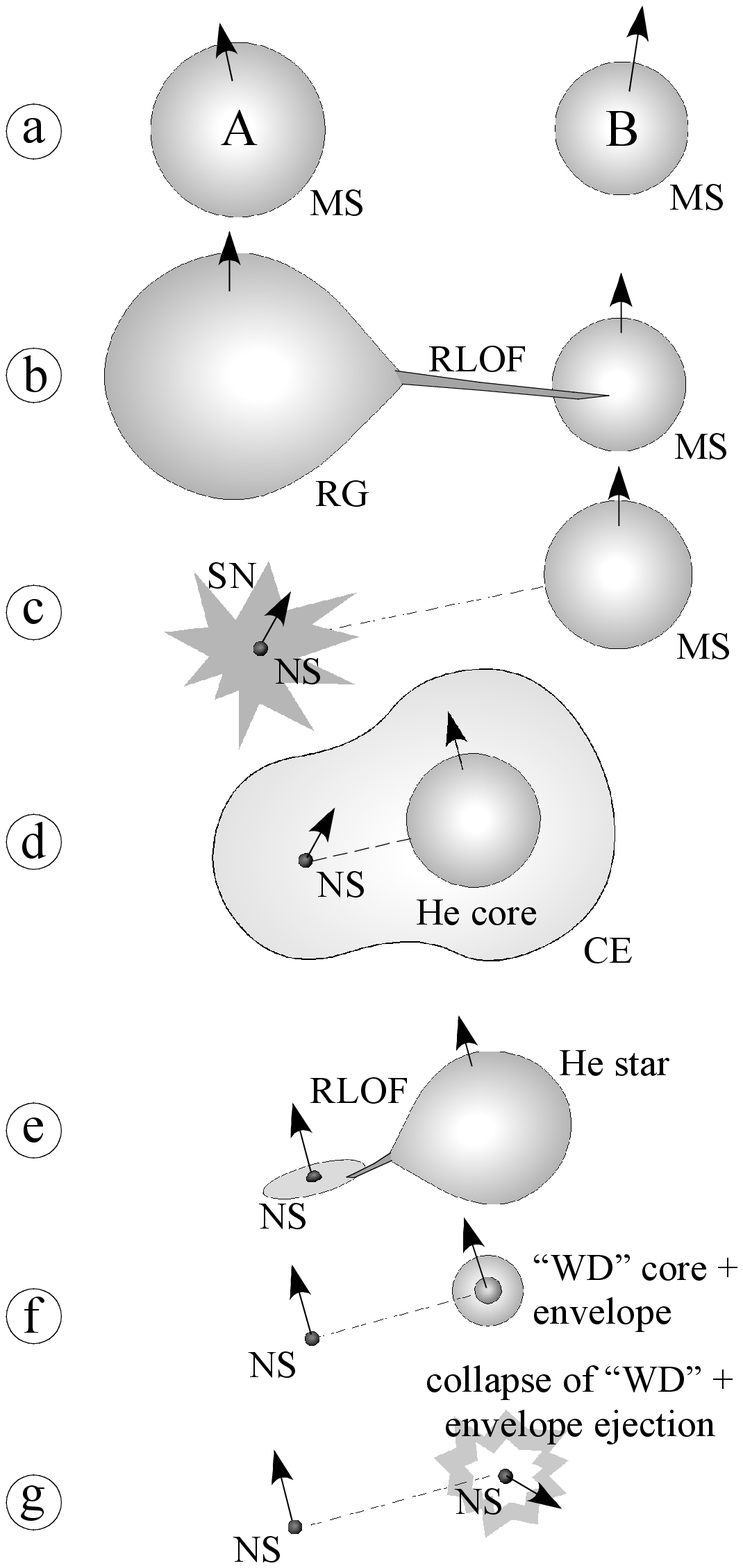}} 
\caption{The evolutionary path of the binary: 
a)	Main sequence phase: Both stars are more massive then 8 solar masses. A is more massive than B. 
b)	Star A becomes a red giant and transfers mass to B via Roche lobe overflow.  Most of the mass of A is lost at this stage. Towards the end of this phase star A contains no more than 4-5 solar masses and possibly significantly less.
c)	Star A collapses ejecting at most three solar masses. During the
collapse the system receives only a mild kick, because of the large mass of star B. 
This kick may change the spin of star A and, to a limited extent, the
orbital spin of the system. 
d)	The neutron star A and the star B form a common envelope phase during which the orbit shrinks significantly and B looses a significant amount of mass.
e)	 Star B continues to shed mass. The accretion spins up the neutron star A and reduces its magnetic field. A becomes a millisecond pulsar and its spin is aligned with the orbital spin. 
f)	Having lost a significant fraction of its mass, B becomes a slowly rotating white dwarf with a tenuous envelope. Both spins are parallel to the orbit's angular momentum.
g)	The white dwarf B collapses to a neutron star emitting neutrinos and ejecting the envelope. The ejected neutrinos give a kick to B changing its spin but causing only an insignificant change to the orbit or to the center of mass motion.
}
\label{fig:evolution}
\end{figure}

The measurement of the proper motion of the system, $\Delta
  v_{{\rm cm}, \perp} = 10 \pm 1$ km/s, confirms with a higher significance that
  pulsar B must have formed in the collapse of a WD-like degenerate core with a tenuous envelope, with the ejection
  of $\sim (0.10 - 0.16)$ M$_{\odot}$. The newborn NS received a very small 
kick  \citep{PiranShavivPRL}. The (near) alignment
  of pulsar A's spin with the orbital plane is consistent with this expectation. 

The slow proper motion of the system can also constrain 
the mass of the stellar progenitor of pulsar A. A significant  ejection of mass during the first SN would have given a large C.M. velocity to the system. 
We find that, at the time 
of its collapse, the progenitor of pulsar A must have been the least massive 
star in the binary. Since it must have been the more massive one  initially, we
can  deduce, from the dynamics of the double NS system, the occurrence
of an early phase of mass transfer in the progenitor stellar binary.

%\item 
The misaligned spin of pulsar B could only be produced during the
  collapse of the WD and the formation of pulsar B. The low kick velocity
  received by pulsar B strongly limits the angular momentum that could be
    given to the nascent NS. This is found to be sufficient to misalign its 
spin axis, provided that Pulsar B inherited little angular momentum from the progenitor WD, 
\textit{and that} the kick was applied at the NS surface or slightly beyond
it, e.g. at the neutrinosphere. 

%\item 
The tilted spin of pulsar B can be viewed as an interesting opportunity to
  test the physical properties of the hot proto-NS. 
Having ruled out all other alternatives, the tilt is found to be consistent with the occurrence of a short-lived convective phase. During this phase the huge neutrino flux is released with a small ($\sim 1\%$) randomly oriented anisotropy. 

%\item 
The required slow rotation of the WD progenitor corresponds to a minimum
spin period of $\sim 3$ hrs. If the core of the He star, where the WD
 was eventually formed, was coupled to the envelope during the $\sim 10^5$ yrs
of RLOF prior to collapse, then an even slower rotation could be
 expected. If, on the other hand, the coupling was not effective the core 
would retain all the {\textit{aligned}} angular 
momentum of the stellar progenitor, eventually forming a fast rotating WD with an  
{\textit{aligned}} spin. In this case, {an off-center kick applied to the
  proto-NS would not have been able to affect the spin direction. 
From this perspective, the tilted spin of pulsar B could be viewed as an
indirect indication of an efficient core-envelope coupling in the
progenitor star.

%\item 
A special role in setting the WD spin could be played by the tidal
 influence of the NS companion, during the $\sim 10^5$ yr long RLOF phase. 
If sufficiently extended convective regions developed in the He-star during
that stage, or if strong magnetic stresses developed across the core-envelope 
boundary, then efficient synchronization of the stellar core might have 
occurred. In this case the spin rate of the hot WD would have been set to the 
$\gtrsim 3$ hr orbital period, meeting the minimal requirement of
eq. \ref{eq:fastest-possible}. 
This very intriguing possibility can be confirmed by detailed evolutionary 
calculations able to follow the He star structure during different nuclear 
burning stages.

To conclude we have found that the progenitor of Pulsar B within the binary system PSR J0737-3039 was a hot  degenerate core with a tenuous envelope of $0.1-0.16 M_\odot$. The envelope was ejected during the collapse and the small kick imparted, most likely by the asymmetry in the escaping neutrinos, gave the pulsar its unique spin axis. Moreover, before the collapse the degenerate core  was slowly rotating.  The overall slow proper motion  of the system also implies that the progenitor of Pulsar A was the less massive star in the system, just before it collapsed. This demonstrates the existence of an early phase of mass transfer within the system.

These conclusions have several interesting implications. 
First, this is  a unique new channel of neutron star formation that was not previously described in the literature. 
Its existence should be considered in any population synthesis calculations that attempt to estimate the rate of neutron star
binary mergers. One should also explore evolutionary tracks that would lead the progenitor star to this stage. 
Given the small mass of the envelope the collapse was not accompanied by any kind of known supernovae. 
It might have produced a very faint transient. Given the great interest in transients nowadays it is worthwhile to try estimating the observed signal and  search 
for such events. Finally the rotational history of the progenitor provides evidence for a sufficient coupling between the core and the envelope at an earlier  stage, possibly resolving the controversy concerning core-envelope coupling. 

This research was supported by an ERC advanced grant (GRBs) and by the  I-CORE 
Program of the Planning and Budgeting Committee and The Israel Science Foundation (grant No 1829/12).

\def\nat{nature}
\def\mnras{MNRAS}
\def\apjl{{Astrophys.\ J.}}
\def\apj{{Astrophys.\ J.}}

\vfill
\eject

\bibliographystyle{apj}
\bibliography{BinaryPulsar-FINAL}
\end{document}